\documentclass[12pt,preprint]{aastex}
\usepackage{epsfig}
\usepackage{amssymb}
\usepackage{graphicx}

\slugcomment{To appear in ApJ Letters}

\shorttitle{A Near-IR  Source  Within the Gap  of FL~Cha}
\shortauthors{Cieza, L.}

\begin{document}

\title{Sparse Aperture Masking Observations of the FL~Cha Pre-transitional Disk}

\author{
Lucas A. Cieza\altaffilmark{1,9}, 
Sylvestre Lacour\altaffilmark{2},
Matthias R. Schreiber\altaffilmark{3,10},
Simon Casassus\altaffilmark{4,10},
Andr\'es Jord\'an\altaffilmark{5,10},
Geoffrey S. Mathews\altaffilmark{1},
H\'ector C\'anovas\altaffilmark{3,10},
Fran\c cois M\'enard\altaffilmark{6,10},
Adam  L. Kraus\altaffilmark{1,11},
Sebasti\'an P\'erez\altaffilmark{4},
Peter Tuthill\altaffilmark{7}, and
Michael J.  Ireland\altaffilmark{8}
}

\altaffiltext{1}{Institute for Astronomy, University of Hawaii at Manoa,  Honolulu, HI 96822, USA}
\altaffiltext{2}{LESIA, CNRS/UMR-8109, Observatoire de Paris, UPMC, Universit\'{e}  Paris Diderot, 5 place Jules Janssen, 92195, Meudon, France}
\altaffiltext{3}{Departamento de Fisica y Astronomia, Universidad de Valpara\'{\i}so, Valpara\'{\i}so, Chile}
\altaffiltext{4}{Departamento de Astronom\'{i}a, Universidad de Chile, Camino del Observatorio 1515, Las Condes, Santiago,  Chile}
\altaffiltext{5}{Departamento de Astronom\'{i}a y Astrof\'{i}sica, Pontificia Universidad Cat\'{o}lica de Chile, 7820436 Macul, Santiago, Chile}
\altaffiltext{6}{UMI-FCA, CNRS / INSU France (UMI 3386)}
\altaffiltext{7}{School of Physics, University of Sydney, NSW 2006, Australia}
\altaffiltext{8}{Department of Physics and Astronomy, Macquarie University, NSW 2109, Australia}
\altaffiltext{9}{ \emph{Sagan} Fellow, lcieza@ifa.hawaii.edu}
\altaffiltext{10}{Millenium Nucleus ``Protoplanetary Disks in ALMA Early Science", Universidad de Chile, Casilla 36-D, Santiago, Chile}
\altaffiltext{11}{ \emph{Hubble} Fellow}

\begin{abstract}
\noindent
We present deep Sparse Aperture Masking  (SAM) observations obtained with the ESO  Very Large Telescope
of the pre-transitional disk object FL Cha (SpT=K8, d=160 pc), the disk of which is known to have a wide optically 
thin gap separating optically thick inner and outer disk components.  
We find non-zero closure phases, indicating a significant flux asymmetry in the K$_{S}$-band emission (e.g., a departure from a single point source detection). 
We also present radiative transfer modeling of the SED of the FL~Cha system
and  find that the gap extends from 0.06$_{-0.01}^{+0.05}$  AU to 8.3$\pm$1.3 AU. 
We demonstrate  that the non-zero closure  phases can be  explained almost equally  well by starlight scattered off  the inner edge of the \emph{outer} 
disk  or by a (sub)stellar companion.
Single-epoch, single-wavelength SAM  observations of transitional disks with large cavities that could become resolved 
should thus be interpreted with caution, taking the disk and its properties into consideration. 
In the context of a binary model,
the signal is most consistent with  a  high-contrast ($\Delta$K$_{S}$ $\sim$4.8 mag) source
at a $\sim$40 mas (6~AU)  projected separation. However, the flux ratio and separation parameters remain highly degenerate 
and a much brighter source ($\Delta$K$_{S}$ $\sim$1 mag) at  15 mas (2.4~AU) can also reproduce the signal.
Second-epoch, multi-wavelength observations are needed to establish the nature of the 
SAM detection in FL Cha.
\end{abstract}


\section{Introduction}

Primordial circumstellar disks with inner cavities or gaps are known as ``transitional" disks and can be identified by their 
peculiar Spectral Energy Distributions (SEDs), which have reduced levels of near- and/or mid-IR excesses with 
respect to the vast majority of Classical T Tauri star  (CTTS) disks. Nevertheless, not all transitional disk SEDs 
look the same and it is clear that not all inner holes and gaps are produced in the same way. Grain-growth, 
photoevaporation, and dynamical interactions with (sub)stellar companions can all result in transitional disk SEDs 
(see Williams $\&$ Cieza, 2011, for a recent review). Distinguishing between the processes that could be 
responsible for the unusual SEDs  of transitional disks requires considerable information: SED shapes, accretion 
rates, disk masses, and multiplicity information (Najita et al. 2007). 

 FL Cha belongs to a subclass of transitional objects known as ``pre-transitional" disks,  which is characterized by a pronounced 
 ``dip" in the mid-IR SED (Espaillat et al. 2007). 
The SEDs of  pre-transitional disks can be reproduced with  models presenting  
a wide ($>$5-10~AU) optically thin gap separating  an optically thick inner disk from an optically thick outer disk.  
The  inner and the outer disk components of pre-transitional objects have  already been resolved by 
long-baseline interferometry observations in the near-IR (e.g., Olofsson et al. 2011; Tatulli et al. 2011) and 
the submillimeter (e.g., Andrews et al. 2011), respectively, confirming the gapped structure.
These gaps are not necessarily empty. Deep polarized intensity images reveal a population of $\mu$m-sized grains within
some of the gaps (Dong et al. 2012) and  the accretion rates onto the stars in pre-transitional systems 
suggest that the inner disks are  continuously replenished with material from the outer disks. 
Since pre-transitional disks tend to lack stellar companions (Pott et al. 2010; Kraus et al. 2011),  their gaps 
are best explained by the dynamical interaction of unseen substellar or planetary-mass objects  
embedded in the disk (Dobson-Robinson $\&$ Salyk, 2012). 
Recent high-contrast observations using the Sparse Aperture Masking (SAM) interferometric technique 
on Keck and the VLT have already identified companion candidates to two
other pre-transitional disks,  T Cha (Hu\'elamo et al. 2011) and LkCa 15 (Kraus $\&$ Ireland, 2012).

FL Cha is a K8  CTTS (Luhman   2007)  in the Chamaeleon I molecular cloud, located at 160 pc (Whittet et al. 1997). 
As part of a program aiming the direct detection of young planets in transitional disks (see also Schreiber et al.  2013),  
we have obtained deep VLT-SAM  observations of FL Cha. 
Our SAM data show non-zero closure phases indicating a second source of near-IR emission besides 
the central star. 
We also present radiative transfer  modeling  of the FL Cha SED, in order to better constrain 
the size of the gap, and discuss the possible nature of the newly identified source. 

\section{SAM observations and data analysis}

\subsection{Observations}\label{obs}

The VLT-SAM observations of  FL Cha  were performed on March 6, 2012, using the  ``7-hole" mask (Tuthill et al. 2010) on the
NAOS-CONICA (NaCo) Adaptive Optics system (Lenzen et al. 2003). 
The mask at the pupil-plane blocks most of the light from target and resamples the primary mirror into a set of smaller sub-apertures 
that form a sparse interferometric array with 21 baselines.  SAM observations allow for exquisite calibration of the point spread function
of the stellar primary and the suppression of speckle noise by the application of interferometric analysis techniques, such as the 
measurement of closure phases (the sum of the phases around any 3 triangle of baselines). 
The SAM technique is sensitive to companions in the $\sim$0.5-5 $\lambda$/D separation range (corresponding to
$\sim$30-300 mas for K$_{S}$-band observations in the VLT) and can reach a contrast limit of $\Delta$K$\sim$7 mag at $\lambda$/D (Kraus \&  Ireland, 2012).

The observing sequence consisted of multiple  ``visits"  (4 observations of 25 frames of 10~s in K$_S$-band) of
FL Cha, alternating with observations of  the stars  FI~Cha,  FK~Cha,  and 2MASS~J11082577-7648315 used as calibrators. 
During the same observing run, we also observed the close binary system RX~J1106.3-7721
(Lafreni{\`e}re et al. 2008) in the L$'$-band to validate  our observing strategy and data reduction technique. 
The observations were reduced using the Paris SAMP pipeline as described by Lacour et al. (2011). 
Both FL Cha and RX~J1106.3-7721 show non-zero closure phases, indicting a departure from a single point-source  detection 
(see Figure~1).  

\subsection{Binary model}\label{MCMC}

In order to constrain the properties of the SAM detections, we first performed 
a Monte Carlo Markov Chain analysis using  a simple 
binary model 
with 3 basic parameters: two positional parameters, either separation and position angle (PA) or $\Delta$RA and  $\Delta$dec, and the
magnitude difference ($\Delta$m).
The $\chi^2$ maps of such models are shown in Figure~1 for both FL Cha and  RX~J1106.3-7721. 
Our procedure is identical to the one used by Schreiber et al. (2013).
We find that the binary parameters for the  RX~J1106.3-7721 system are  \emph{very} well constrained by our observations 
and robust to the choice of priors (see Figure 2). 
The flux asymmetry signal in the closure phases of FL Cha is much  weaker 
and
 results in much larger uncertainties in the model-derived parameters. 
 While the PA is relatively well constrained,
$\Delta$m and separation remain degenerate and highly dependent on the choice of prior distributions. 
This is illustrated in Figure 2, where we show the posteriori probability distributions for two different 
sets of priors:
1) uniform distributions for $\Delta$RA, $\Delta$dec, and $\Delta$m and 2) uniform distributions for PA, $\Delta$m, and
the logarithm of the separation.
The data favors a $\Delta$m $\sim$4.8 mag source, but there is a long tail in the probability distribution
extending to lower $\Delta$m values. 
The posteriori distribution of the separation could be bimodal or unimodal  depending on the adopted prior. 
The degeneracy  between the $\Delta$m and separation parameters is a known problem for small separations ($\lesssim$ $\lambda$/D)
in aperture masking observations (Pravdo et al. 2006,  Martinache et al.  2009)  and is clearly seen in the joint distributions shown in Figure~2.  
The complex $\chi^2$ surfaces in the joint distributions  are consistent with two families of solutions: 
a relative bright source ($\Delta$K $\sim$1-3 mag) at  $\sim$15  mas (2.4~AU) and
a much fainter one ($\Delta$K $\sim$4-5 mag) at $\sim$30-40 mas (5.0-6.5~AU).
The relative probability of these two solutions strongly depends on the choice of prior and can not be
unambiguously estimated with the available data.
Future H-band observations should 
provide the additional resolution needed  to break the $\Delta$m-separation degeneracy and solve 
the two-solution ambiguity 
in the context of the binary model
(see Schreiber et al. 2013).

\section{Disk model}\label{disk-model}

Espaillat et al. (2011) successfully reproduced the optical to 38 $\mu$m SED of
FL Cha adopting a simple model consisting of two vertical walls: one inner wall at the dust
sublimation distance of 0.04 AU that is responsible for the near-IR excess and an
outer wall at 15 AU that reproduces  the observed mid-IR excess.  
Here we use a more physical model in order to provide further constraints on the size and location
of the gap in the disk.  We  include photometry data at longer wavelengths (70 and 870 $\mu$m), 
which are sensitive to the properties of the outer disk. 

\subsection{Spectral energy distribution} 

We constructed the FL Cha SED  from the sources listed in Table 1.
We have also obtained 870 $\mu$m photometry  with the Atacama Pathfinder Experiment (APEX) 
using the LABOCA camera (Siringo et al. 2009).  The observations were executed on July 24, 2012
using the ``wobbler on-off" mode  and reduced with the standard bolometer array data analysis package 
BoA\footnote{http://www.apex-telescope.org/bolometer/laboca/boa/}.
FL Cha was detected with a signal to noise ratio of 12. 
The SED of FL~Cha is also well sampled by a \emph{Spitzer}-IRS spectrum covering the 5.2 to 38 $\mu$m
region (Astronomical Observation Request $\#$12696320).
The 0.44 to 870 $\mu$m  SED is shown in Figure~3. The optical and near-IR 
wavelengths were corrected by extinction, adopting  A$_{V}$ = 3.14 mag  (calculated from
the R$_C$-I$_C$  color excess)  and the extinction relations
listed in Cieza et al. (2007). 
The median SED of CTTSs from Furlan et al. (2006) is shown for comparison. 
The sharp ``dip" in the FL Cha SED around 15 $\mu$m indicates the presence of a wide gap 
in the disk. 

\subsection{Radiative transfer modeling}

We model the observed SED using the Monte Carlo radiative transfer code 
MCFOST (Pinte et al. 2006).
We parameterize the structure of the FL  Cha disk with two independent 
components, an inner and an outer disk. Each component is described by
 the following parameters:
 the inner and outer radii  (R$_{in}$ and  R$_{out}$, respectively) and the index
 $\gamma$ for the surface density profile
 ($\Sigma$(r) = $\Sigma_{10}$(r/10 AU)$^{\gamma}$).
The scale height as a function of radius is  given by $H(r)$ = H$_{10}$(r/10 AU)$^{\psi}$.
The grain size distribution has the form  d$n$(a)$~\propto$~a$^{p}$d$a$, 
between the minimum (a$_{min}$) and maximum (a$_{max}$) grain sizes.
The width of the gap in the disk is simply given by: $R_{outer, in}$ - $R_{inner, out}$. 
The  parameters a$_{min}$, a$_{max}$,  $p$, $\psi$, $\gamma$, and R$_{outer, out}$, and
the disk inclination  are not easily constrained by the available data. We thus fix them to more 
or less ``standard"  values, which are listed in Table 2: 
For the dust composition, we follow Espaillat et al. 
(2011) and adopt a 40$\%$/60$\%$ mixture of amorphous and
crystalline silicates. 

We adopt the fitting procedure described by
Mathews et al. (2012), which uses the Levenberg-Marquardt $\chi^2$
minimization algorithm to calculate the numerical 
gradients of the $\chi^2$ function and determine the next point in
the parameter space to be sampled until the algorithm
converges to a $\chi^2$ minimum. 
We ran the search algorithm 10 times using different starting values 
to better sample the parameter space.  
Each of the runs results in a set of best-fit parameters, the distribution of which
can be used to calculate the mean and associated uncertainty.
 
The results of the 10 runs are listed in Table~2 and the structure of our best-fit model is shown in Figure~3.
We find  that the  near-IR excess is best
reproduced by a very narrow ring at the dust sublimation radius, extending from 
0.04 to  0.06 AU, in agreement with the modeling results by Espaillat et al (2011).
We also find that inner disks wider than $\sim$0.1 AU significantly  overproduce the observed near-IR 
fluxes and that the dip characteristic of pre-transitional disk SEDs  disappears for inner disks wider than $\sim$3 AU 
(for the given $R_{outer, in}$ value, resulting in gaps narrower than $\sim$5 AU). 
In our model, the inner edge of the outer disk is located at  8.3$\pm$1.3 AU from the star. This distance 
is a factor of $\sim$2 smaller than that estimated by Espaillat et al. The source of the discrepancy 
is unclear, but since our models share the same stellar parameters, it is likely to be related to the different  parametrizations used 
for the disk structure (two vertical walls versus a full 3-D disk model) and the different grain size
distributions adopted.  
Resolved submillimeter images with ALMA should be able to directly measure the cavity size and
settle the discrepancy. 
Our SED model also constrains the properties of the outer disk. In particular, we 
find the FL Cha disk is relatively ``typical", with   a mass of 7.5 M$_{JUP}$ and 
a scale height of  $\sim$1 AU at a radius of 10 AU.

\section{Possible interpretations of the SAM detection}

\subsection{Background contamination}
 
The probability $P(\Theta,m)$ for an unrelated source to be
located within a certain angular distance $\Theta$ from a particular
target is given by
$P(\Theta,m) = 1 - e^{-\pi\rho(m)\Theta^2}$, 
where $\rho(m)$ is the cumulative surface density of background
sources down to a limiting magnitude $m$ (Brandner et al. 2000). 
FL Cha is 9.11  mag in K$_{S}$-band. Therefore, 
the peak in the  $\Delta$K$_{S}$ probability distribution   
correspond to an apparent  K${_S}-$band  magnitude of $\sim$13.9.
Since there are 7482 stars brighter than K$_{S}$ = 13.9 mag within 
a 1 deg radius of FL Cha in the $2MASS$ catalog, the probability of a background
source at  a $\lesssim$0.050$\arcsec$ separation
is of the order of $5\times10^{-6}$. 
Background contamination can thus be discarded
as a likely explanation for our SAM detection.

\subsection{A  stellar companion}\label{companion}

In Section~\ref{MCMC},  we  found the SAM data is consistent with both
a relative bright source ($\Delta$K$ \sim$1-3 mag)  at  a projected separation 
of $\sim$2.4 AU or a much fainter one ($\Delta$K $\sim$4-5 mag) at $\sim$5.0-6.5 AU. 
Evolutionary tracks of young low-mass objects are \emph{very} uncertain, 
but can be used to address the nature of  the putative companion and try to 
distinguish between a low-mass star, a brown dwarf, or a planet. 
According to the evolutionary tracks by Siess et al. (2000), FL Cha is  a 0.6 M$_\odot$ star and
the former solution corresponds to a stellar companion with a mass in the $\lesssim$0.1-0.3 M$_{\odot}$
range. 
In the recent  radial velocity (RV) monitoring study of  Chamaeleon I objects performed by 
Nguyen et al. (2012),  FL Cha showed a constant RV  (16.9$\pm$1.1 km/s), close to the typical values in  
the region ($\sim$15.3$\pm$2 km/s),  over the 1 month baseline of the study.
While the precision and time baseline of the measurements are clearly not enough to rule out 
most stellar binaries,  the RV measurements disfavor  solutions  with stellar companions at  $\sim$2.4 AU.
As a reference, a 0.1 M$_\odot$ companion to a  0.6 M$_\odot$ star with  circular edge-on orbit  and a 2.4 AU semi-major axis 
has a period of 4.46 years and a velocity amplitude of $\pm$2.5 km/sec.

\subsection{A  brown dwarf or a protoplanet}\label{bd}

 We now consider the nature of the source if $\Delta$K$_S$ is close to the $\sim$4.8 mag peak shown in the 
 probability distributions from  Figure 2.
This  peak corresponds to
%
an absolute  magnitude of  $\sim$7.9  at 160 pc.
This renders our source a factor of 3 brighter than the protoplanet candidate
identified by Kraus $\&$ Ireland (2012) within the gap of the LkCa 15 disk,
and the ``hot start"  models by Chabrier et al.  (2000) assign it a mass of 
$\sim$15-20 M$_{JUP}$, for an age of 1 Myr. Taken at face value, this would
place the object at the bottom of the brown dwarf mass function. 
However, since the source seems to be located inside the gap of an 
accreting transitional disk, the inner disk of which is mostly depleted, it
is reasonable to expect material from the outer disk  to 
flow across the gap onto the inner disk and then the star. 
Under such circumstances, the low-mass object inside the gap  should accrete
most of the material being transported across the outer disk  (Lubow $\&$ D'Angelo, 2006), 
resulting in significant accretion luminosity. 
SAM observations in H- and L-band would provide near-IR colors 
and help establishing whether  the FL Cha detection is consistent with a  brown dwarf or an actively accreting 
protoplanet surrounded  by a disk, as seems to be the case in the  LkCa~15 system.

 \subsection{Thermal emission or starlight scattered off the disk}

Our SAM  detection is inconsistent with  direct thermal emission
from the \emph{inner} disk. Our radiative transfer model suggests  that the inner disk is 
$\lesssim$1 mas in diameter and the thermal contribution from the outer disk to the observed 
K-band flux is $<$0.01$\%$.
However,  starlight scattered off the inner edge of the  \emph{outer} disk
can in principle produce the observed signal. 
If the disk is highly inclined, the brightness of the projected rim would be 
asymmetric and could result in non-zero closure phases.  
To test  this hypothesis, we generated a 2.2  $\mu$m ray traced image of 
the best-fit MCFOST model described in Section~\ref{disk-model} and calculated 
the closure phases of the resulting image as seen at different position
angles and inclinations (Figure~\ref{fig:disk}).
We find that a disk inclined by $\sim$60 $\deg$ from face-on with a position angle\footnote{The position angle is the 
direction defined  by the intersection between the plane of the disk and the plane of the sky.} of $\sim$150 deg 
results in closure phases that fit the  SAM data almost as well as the best-fit binary model described in Section~\ref{MCMC} does
($\chi^2$ $\sim$140 vs. $\sim$130).
The difference in the $\chi^2$ results is mostly due to the fact that  the disk model produces slightly smaller phase values 
compared to both the binary model and the SAM data.  
However, we emphasize that the 2.2 $\mu$m model image depends on disk properties that are poorly constrained, such 
as the grain size distribution, the dust composition,  and the detailed structure of both the inner and the outer disk. 
These disks parameters could be varied to improve the fit further. Pending more constraints (i.e., resolved images), the current model 
suffices to show that a disk (which also fits the SED) can produce the necessary closure phase signal.
This demonstrates that \textit{the circumstellar disk must be taken into consideration
when interpreting SAM data of transitional disks with inner cavities that are  large enough to become resolved.} 
High-resolution submillimeter imaging with ALMA would reveal the exact orientation and  size 
of the cavity in the FL Cha disk.  
Such imaging should 
establish whether our SAM detection is in fact consistent with starlight scattered off the disk or,  
on the contrary,  if the source is located well inside the hole, as in the case of 
the LkCa 15 system. 
Measuring  the K$_S$-H  and K$_S$-L colors of the source would also help
testing the scattered light  hypothesis.
Looking for orbital motion from multi-epoch SAM observations should provide 
 the ultimate test to distinguish between a companion and the
 scattered light scenario. 
 
\section{Summary and Conclusions}

From  VLT-SAM observations,  we have identified 
a near-IR flux asymmetry in the pre-transitional object FL Cha. 
By modeling its SED, we find that the gap in its disks extends from 0.06$_{-0.01}^{+0.05}$  AU to 8.3$\pm$1.3 AU.
We have considered several potential possibilities for the nature of the  source: 
a low-mass star, a brown dwarf,  a protoplanet,  thermal emission or starlight scattered off  the disk, or a background object. 
Only  direct thermal emission from the inner  disk and background contamination can be ruled out. 
We find that light scattered off the inner edge of the \emph{outer} disk can result in closure phases that fit the SAM data 
almost as well as binary models do.
Single-epoch, single-filter SAM observations of transitional disks should thus be interpreted very cautiously, taking the
disk into consideration.  
In the context of a binary model, the closure phases are most consistent with  a  $\Delta$K$_{S}$ $\sim$4.8 mag  source
at  a 6~AU  projected separation,  but a much brighter one ($\Delta$K$_{S}$ $\sim$1 mag) 
at  2.4 AU can also reproduce the signal.
H- and L-band SAM observations of  FL Cha are highly desirable to 
1) test for orbital motion,
2) break the current degeneracy between $\Delta$mag  and separation in the binary model, and
3) provide near-IR colors to help distinguishing  between the possible explanations. 
Resolved submillimeter  images with ALMA  are needed to establish the orientation
of the system and provide a  \emph{direct} measurement of the inner cavity size to test the scattered light
hypothesis  and  to better constrain  the properties of the outer disk. 

\acknowledgments
L.A.C was supported by NASA through
the \emph{Sagan} Fellowship Program.
M.R.S.,  A.J., S.C., and F.M. acknowledge
support from the  Millenium Science Initiative, Chilean Ministry of Economy, Nucleus P10-022-F.
G.S.M.  acknowledges NASA/JPL funding support
through grant RSA-1369686. 
A.L.K. was supported by NASA through the \emph{Hubble}
Fellowship program.

\newpage

\begin{deluxetable}{ccrlrr}
\footnotesize
\tablecaption{ FL Cha photometry data}
\tablehead{\colhead{Wavelength}&\colhead{Flux}& \colhead{Flux}& \colhead{Error$^a$}&\colhead{Telescope}&\colhead{Reference$^b$}\\
                    \colhead{($\mu$m)}&\colhead{(mJy)}&\colhead{(mag)} &\colhead{(mJy)}& \colhead{}&\colhead{} }
\startdata
0.44   &    1.74E--01  &   18.44 & 30$\%$ & ground based  & 1 \\
 0.55  &     1.16E+00 &   16.28&  30$\%$ &   ground based  & 1 \\
 0.65  &       3.19E+00&     14.96  &    20$\%$ &  ground based &   1   \\
 0.80   &     1.10E+01 &        13.41   &  20$\%$ &   ground based & 1   \\
 1.25    &   5.45E+01      &   11.73 &  15$\%$    & 2MASS  &   2 \\
  1.66  &    1.17E+02&         9.90&   15$\%$       &   2MASS & 2 \\
  2.20  &    1.51E+02  &       9.11    &    15$\%$   &  2MASS & 2 \\ 
  3.6   &    9.66E+01  &  8.66        &   10$\%$       & \emph{Spitzer} & 3 \\
  4.5  &     8.20E+01   &  8.35       &    10$\%$    & \emph{Spitzer} & 3 \\
  5.8  &     5.60E+01&    8.28      &   10$\%$     & \emph{Spitzer} & 3\\
  8.0  &     3.95E+01 &    8.02      &   10$\%$      & \emph{Spitzer} & 3 \\
  24   &  1.01E+02         &    4.67          & 10$\%$  & \emph{Spitzer}  & 3 \\
  70   &  3.01E+02        &   \nodata   & 15$\%$  &  \emph{Spitzer}   &  3 \\
  870 &  3.00E+01    &   \nodata   &15$\%$  &           APEX          & 4
\enddata
\tablecomments{
$^a$the optical and near-IR uncertainties are dominated by the extinction corrections.
$^b$References:  (1)   Gauvin $\&$ Strom (1992);  (2) Skrutskie et al.  (2006);  (3) \emph{Spitzer}'s  Gould Belt Catalog. 
The 3.6--8.0 $\mu$m photometry  have already been published by  Cieza et al. (2009).  
The 24 and 70 $\mu$m data have not been published before;   (4) this work. 
}
\end{deluxetable}

\begin{deluxetable}{lrr}
\footnotesize
\tablecaption{Disk and Stellar Parameters }
\tablehead{\colhead{Parameter}&\colhead{value}&\colhead{error}}
\startdata
Stellar parameters && \\
\hline
Stellar T$_{eff}$         [K]                       & 3850    &  fixed     \\
Stellar Luminosity     [L$_{\odot}$]     & 0.4      & fixed \\
Stellar mass   [M$_{\odot}$]   &   0.6    &  fixed \\
Distance        [pc]                           &   160   &  fixed \\
\hline
Inner and outer disk parameters  && \\
\hline
Inclination     [deg]                         &  60      & fixed  \\
Grain size distribution slope, p   &  $-$3.5   &   fixed \\
a$_{min}$  [$\mu$m ] &  0.005     &  fixed \\
a$_{max}$ [$\mu$m] &  3900         &  fixed \\
Surface density exponent, $\gamma$ &  -1   &  fixed \\
Flaring exponent,  $\psi$ &  1.1   &  fixed \\ 
\hline
Inner disk parameters & &   \\
\hline
Scale height at 10 AU  H$_{10, inner}$  [AU]  &  0.16 & 0.05 \\
 Mass$_{disk, inner}$ [M$_{JUP}$]\tablenotemark{1}   &  5$\times$10$^{-4}$& 5$\times$10$^{-4}$  \\  
R$_{inner, in}$        [AU]  &      0.04  &       0.01    \\
R$_{inner, out}$ [AU]   &          0.06 &        0.05        \\
\hline
Outer disk parameters  & & \\
\hline
Scale height at 10 AU  H$_{10, outer}$  [AU]  &  1.2 &  0.7 \\ 
Mass$_{disk, outer}$ [M$_{JUP}$]\tablenotemark{1}  &  7.5 & 1.0 \\  
R$_{outer, in}$        [AU]  &      8.3 &       1.3    \\
R$_{outer, out}$   [AU]  &  100  & fixed \\
\enddata
\tablenotetext{1}{
Assumes a gas to dust mass ratio of 100.
}
\end{deluxetable}

\begin{figure}
\includegraphics[width=15cm, trim = 0mm 0mm 0mm 0mm, clip]{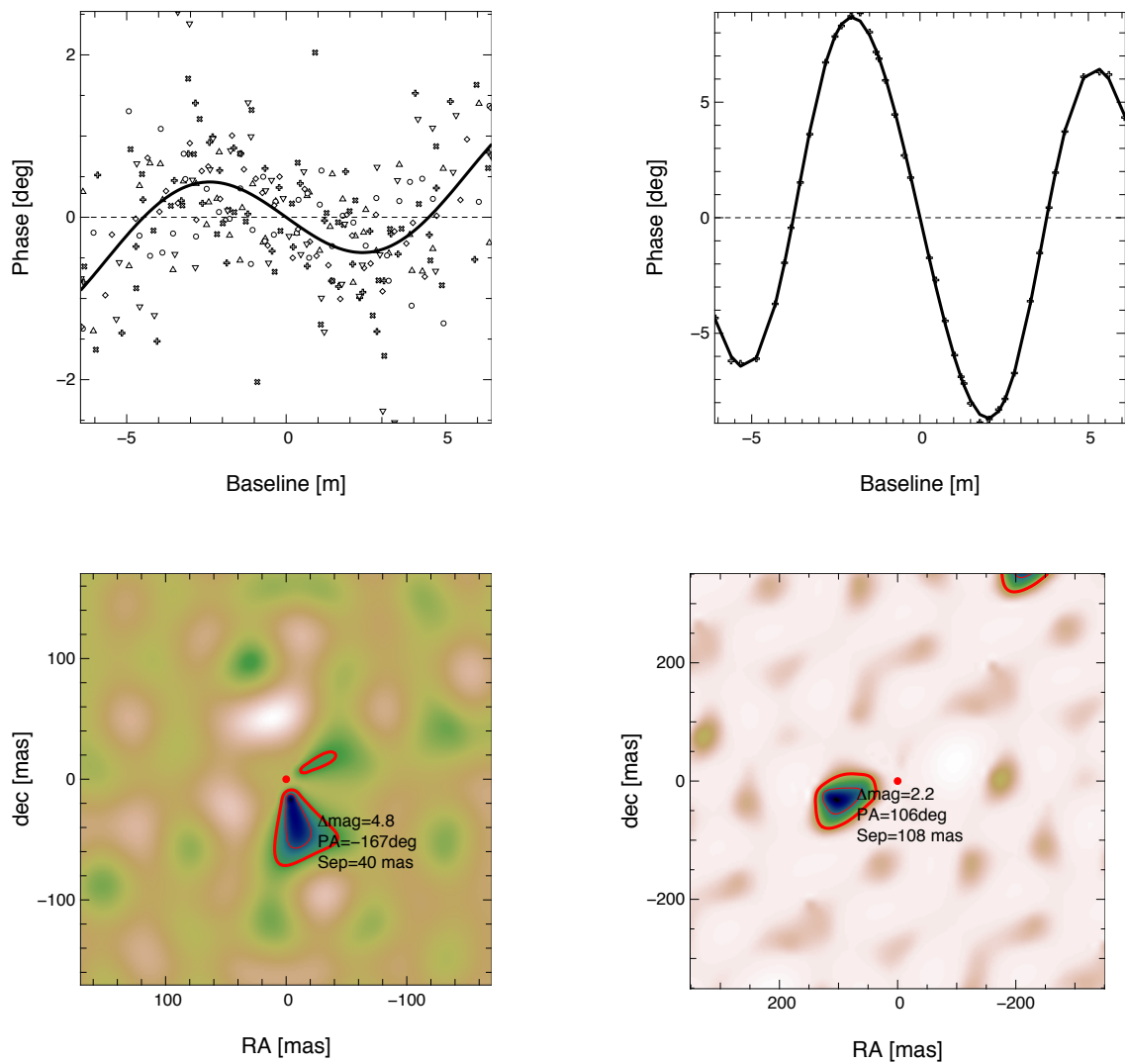}
\caption{ \small 
\textbf{Top Panels:} Closure phase as a function of baseline from our VLT-NACO/SAM
data indicating significant flux asymmetries in both FL Cha (left) and  RX~J1106.3-7721 (right).
\textbf{Lower Panels:}  $\chi^2$ maps resulting from fitting a
binary  model to the closure phases. The 5-$\sigma$ contours indicate the location of the best-fit companions.
}
\label{fig:detection}
\end{figure}

\begin{figure}
\includegraphics[angle=0, width=15.cm, trim=0mm  0mm 0mm 0mm]{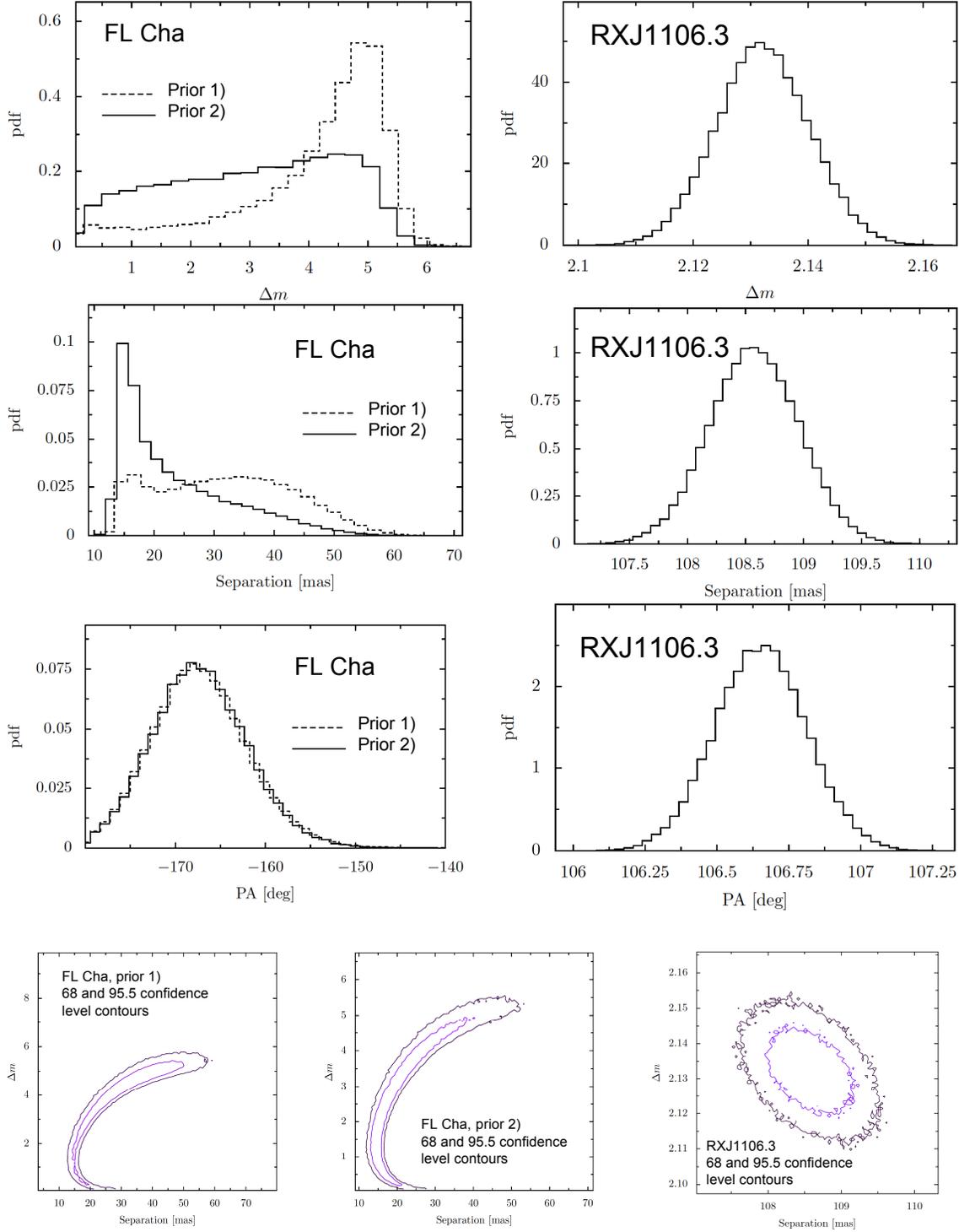}
\caption{ \small \textbf{Top Rows:} the posteriori distributions of the 3 parameters in the binary models ($\Delta$mag, 
separation, and PA) for FL Cha (left, using two sets of priors as described in Section~\ref{MCMC}) and RX~J1106.3-7721 (right). 
\textbf{Bottom Row:} The joint distributions of separation and $\Delta$mag. For FL Cha,
 these two quantities  remain degenerate and highly dependent on the choice of priors.  
}
\label{fig:MCMC}
\end{figure}

\begin{figure}
\includegraphics[angle=0, width=16.cm, trim=0mm  0mm 0mm 30mm]{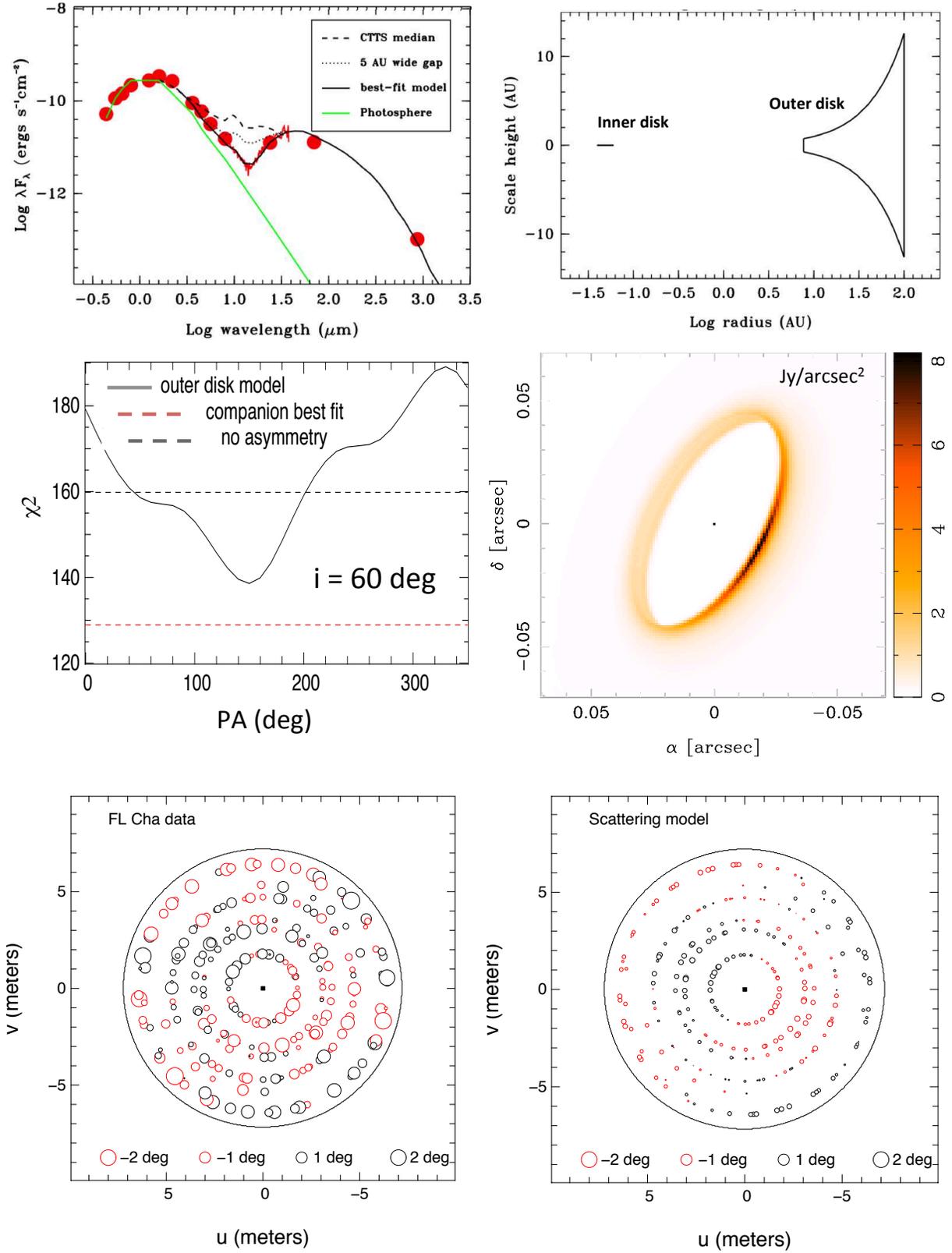}
\caption{ \small  \textbf{Top Panel:} The extinction-corrected SED of FL Cha showing 
a pronounced ``dip" at mid-IR wavelengths, the defining feature of pre-transitional disks (left) and 
the disk structure of our best-fit SED model (right). 
\textbf{Middle Panel:} Quality of fit to the SAM data for our best-fit disk model as a function of disk PA, compared to
the best-fit binary model and a single point source (left).  K$_S$-band ray traced image of the best-fit disk model (right). 
\textbf{Bottom Panel:} UV coverage of the SAM observations on FL Cha indicating the sign and value of the phases (left). The 
phases resulting from the best-fit disk model adopting the same UV coverage as in the real data (right). 
}
\label{fig:disk}
\end{figure}

\end{document}